\documentstyle[aps,twocolumn,epsfig]{revtex}

\begin{document}

\newcommand{\sgn}{\; \hbox{sgn}\,}
\newcommand{\op}[1]{{\rm \hat{#1}}}
\newcommand{\bra}[1]{\langle #1|}
\newcommand{\ket}[1]{|#1\rangle}
\newcommand{\braket}[2]{\langle #1|#2\rangle}

\title{Anomalous diffusion and dynamical localization in a parabolic map}
\author{Toma\v z Prosen and Marko \v Znidari\v c}
\address{Physics Department, Faculty of Mathematics and Physics, 
University of Ljubljana, Jadranska 19, 1111 Ljubljana, Slovenia}
\date{\today}
\draft
\maketitle
\begin{abstract}
We study numerically classical and quantum dynamics of a piecewise parabolic area preserving 
map on a cylinder which emerges from the bounce map of elongated triangular billiards. 
The classical map exhibits anomalous diffusion. Quantization of the same map  
results in a system with dynamical localization and pure point spectrum. 
\end{abstract}
\pacs{PACS numbers: 05.45.Pq, 05.45.Mt} 

As the Schr\" odinger equation is {\em linear}, it cannot posses 
exponential instability in time as do the classical equations of motion. 
Quantum time evolution is dynamically stable \cite{Casati86} and all Lyapunov 
exponents are strictly zero. This does not contradict the correspondence principle as 
the quantum evolution still mimics classical exponential instability up to so-called
{\em log-time} $\log{1/\hbar}$. For larger times, the classical evolution develops 
details on scales smaller than $\hbar$ and the strict correspondence is lost. 
In some quantum systems there is another time scale, connected with the phenomena 
of {\em dynamical localization}. This is purely quantum interference effect and 
results in a quantum suppression of classical diffusion. For times smaller than 
this so-called {\em break time} $t<t_{\rm B}$, quantum evolution follows the classical 
diffusion. Afterwards, the quantum system ``notices'' it's discrete spectrum and the 
diffusion stops, resulting in a localized state. Since the break time $t_{\rm B}$ is 
usually much larger than the log-time, exponential instability may {\rm not be} 
relevant for dynamical localization, although all known examples of dynamical 
localization, from kicked rotors and maps\cite{ChirikovCasati,Izrailev} to quantum 
billiards\cite{BorgCasLi}, take place in classically chaotic systems. 
\par
There are several reasons for studying systems without exponential instability.
Most important is to gain insight into the old question, namely what are the 
necessary and sufficient properties of a system in order to display certain 
statistical properties. This question can be addressed in classical as well as in 
quantum mechanics. In this Letter we study
a simple dynamical model with purely parabolic, non-chaotic and non-integrable, 
classical time evolution. By means of numerical experiments and heuristic 
arguments we show that the model exhibits anomalous classical diffusion 
whose mechanism is essentially different from the one that is 
commonly observed in softly chaotic, i.e. KAM, systems. 
Furthermore, we show that the quantization of the same model displays dynamical localization, 
supporting a provocative claim that {\em dynamical localization} can 
exist {\em without classical chaos} and {\em without normal classical diffusion}.
\par
Triangular billiards are among the simplest classical systems without exponential 
instability. It is worth mentioning that a triangular billiard is equivalent to the 
motion of three particles on a ring (provided all three angles are smaller than $\pi/2$) 
with their masses connected to the angles of a triangle \cite{Casati&Prosen99}. 
Here we consider a special asymptotic case of an elongated triangular billiard
with one angle $\gamma$ being very small \cite{Casati&Prosen00}, where 
it was shown that in the limit $\gamma\to 0$ the bounce map, 
relating two successive collisions with the short sides of a triangle read 
\begin{eqnarray}
v_{n+1} &=& v_n+2(u_n-[u_n]-\mu (-1)^{[u_n]}) \nonumber \\
u_{n+1} &=& u_n-2v_{n+1} \pmod{2},
\label{uvmap}
\end{eqnarray}            
where $u,v$ are suitable position-momentum Birkhoff coordinates, $\mu$ is a 
parameter measuring the asymmetry of two non-small angles, and $[u_n]$
is the nearest integer to $u_n$. Definition range for variables $(u,v)$ is a 
cylinder. Furthermore, if compactified onto a torus 
$(u\hbox{ (mod }2),v\hbox{ (mod }1))$, the map (\ref{uvmap}) can be rewritten in 
new variables $x,y\in [-1,1)$, with 
$y_n=2(-1)^n(u_n+v_n-\frac{1}{2})$ and $x_n=(-1)^n(u_n-\frac{1}{2})$,
\begin{eqnarray}
y_{n+1}&=&y_n+4\mu \sgn x_n \quad (\hbox{mod }2) \nonumber \\
x_{n+1}&=&x_n+y_{n+1} \qquad \quad (\hbox{mod }2),
\label{xymap}
\end{eqnarray} 
where $\sgn x_n$ denotes the sign $\pm 1$ of $x_n$. Both maps, the map 
(\ref{uvmap}) 
and the so called triangle map (\ref{xymap}) are area preserving and (piece-wise) 
parabolic, therefore they do not exhibit exponential instability. 
We stress once again that the maps (\ref{uvmap}) and (\ref{xymap}) 
are equivalent only if both are considered on the 2-torus. 
In the present paper we will focus on the classical and quantum properties of the 
map (\ref{uvmap}) on the cylinder. Let us first discuss its classical properties. 
\par
If considered on the torus (\ref{xymap}), three different cases of parameters must 
be distinguished: (i) If the parameter $\mu=k/2l$ and the initial $y_0=2m/l$ are both 
{\em rational} ($x_0$ can be arbitrary irrational) then the dynamics can be reduced to 
a permutation on a finite integer lattice $l\times l$, hence all orbits are 
{\em periodic} with period which divides $l^2$.
(ii) The parameter $\mu=k/2l$ is {\em rational}, but $x_0$ and $y_0$ are 
{\em irrational}. In this case, the dynamics is {\em pseudointegrable} and the 
motion appears to be {\em ergodic} on the invariant curve which is a set of $l$ {\em helices}
$(y - y_0) l \hbox{ (mod }2)=0$. 
(iii) If the parameter $\mu$ as well as initial coordinates $x_0,y_0$ are
{\em irrational} then the motion appears to be {\em ergodic} on the whole phase space. 
However, the exploration of the phase space is very slow, namely it was found 
numerically that the number of different values of $y_t$ visited up to time 
$t$ grows as $\sim \log t$, i.e. {\em weak ergodicity} \cite{Kaplan&Heller98}. 
\par
If we consider the same three cases in the original 
variables $u$ and $v$, that is the map (\ref{uvmap}) on the cylinder, the case (i) 
results in a ballistic transport as generic periodic orbits of the triangle map 
now result in ballistic orbits.
In pseudointegrable case (ii) one may focus on the motion along infinite invariant 
helices ($(u + v)l \hbox{ (mod }1) = {\rm const}$).
Numerical evidence suggests strongly that the motion along the helix 
exhibits anomalous diffusion $(v_n-v_0)^2 \sim n^\alpha$, with exponent 
$1 < \alpha < 2$ which depends on parameters $\mu,u_0,v_0$.
In weakly ergodic case (iii), we have `logarithmically slow' diffusion transverse 
to the helices due to weak ergodicity and `faster than normal' diffusion along the 
helix. As a result we find numerically significant power-law anomalous diffusion in 
momentum variable $(v_n-v_0)^2 \sim n^\alpha$.
Therefore, the transport properties of the triangle map (\ref{uvmap}) in cases 
(ii) and (iii) are quite similar, hence the irrationality of $\mu$ does not seem to 
play very important role, apart from determining the dimensionality of the 
invariant surface. From now on, we set in all numerical experiments 
$\mu=(\frac{1}{2}(\sqrt{5}-1)-e^{-1})/8$ (same as in \cite{Casati&Prosen00}), but 
note that results are practically independent of this particular value.       
\par
\begin{figure}
\centerline{\psfig{figure=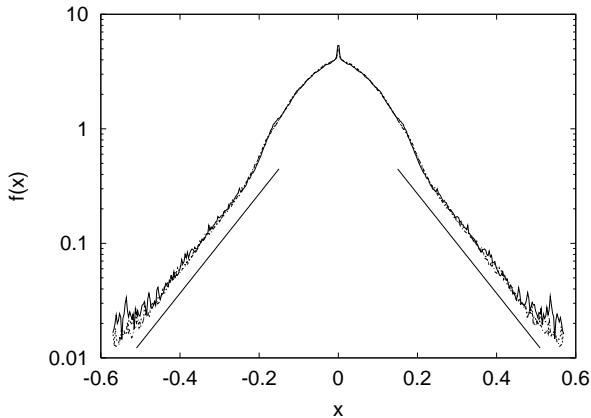,width=2.25in,angle=-90}}
\caption{Scaling function $f(v/t^{\alpha/2})$. Scaled density distributions for three different times $t=6000$, 
$t=12000$ and $t=18000$, full, dashed and dotted curves, respectively, are shown and all fall on the same curve. 
Averaging over $20\cdot 10^6$ initial conditions is performed. Asymptotic solid line is exponential function.}
\label{fig:scaling}
\end{figure}
\begin{figure}
\centerline{\psfig{figure=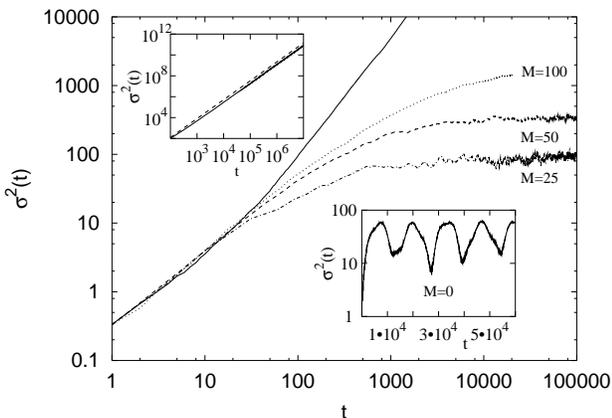,width=2.25in,angle=-90}}
\caption{Anomalous diffusion in a classical map (\ref{uvmap}) (solid curve), and 
quantum cases for $M=100,50,25$ (dot-dashed curves), and $M=0$ (lower inset).
In the upper inset we show classical spreading (solid curve) and $t^{\alpha}$ with 
best-fitting exponent $\alpha=1.81$ (dashed curve).}
\label{fig:localization}
\end{figure}

If we denote by $\rho(v,t)$ the density of a swarm of trajectories along the 
momentum axis as a function of time $t$, $\int{\! \rho dv}=1$, then the dispersion 
$\sigma^2(t)=\int{v^2 \rho(v,t)dv} \propto t^{\alpha}$ grows faster than linear and
defines exponent $\alpha$ (see the inset in fig. \ref{fig:localization}). 
For our map (\ref{uvmap}) the exponent of anomalous diffusion is approximately 
$\alpha=1.81\pm 0.01$ and only weakly depends on $\mu$.
In addition to that, as we show in fig.\ref{fig:scaling}, the momentum-density 
starting from initial $\rho(v,0) = \delta(v)$ satisfies a scaling law 
$\rho(v,t) = t^{-\alpha/2} f(v/t^{\alpha/2})$
where the scaling function $f(x)$ has asymptotically ($|x|\to \infty$) exponential 
tails. Anomalous diffusion is usually explained in terms of L\' evy walks which 
emerge as a consequence of some hierarchical structures in phase space that exhibit scaling. 
For smooth Hamiltonian systems, these structures are stable KAM islands with sticky boundaries. 
Orbit wanders throughout the chaotic sea, but now and then it gets trapped around one of 
those islands, causing temporary ballistic motion. However, it seems obvious that
successive ballistic flights are {\em uncorrelated} as the orbit
spends the meantime in the chaotic sea with positive Lyapunov exponent,
justifying a random walk approximation, which gives relation between the exponent of 
anomalous diffusion and the distribution of flight lengths for 
{\em individual} ballistic periods \cite{Levy}. To our knowledge, all so far studied 
cases of anomalous diffusion involved systems with Lyapunov chaos, except recently 
studied ``square-ring'' billiard \cite{Artuso00}. For systems without exponential 
instability the connection between L\' evy flights and the exponent of anomalous 
diffusion cannot be so simple. This is clearly confirmed by our experiments. 
The distribution of individual flight lengths has been found to fall as 
$(\Delta v)^{-\beta}$ with 
$\beta \approx 2.9$, whether according to the random walk approximation this exponent 
should be $\beta = 4-\alpha \approx 2.2$. The reason for the discrepancy are strong 
correlations between consecutive flights. We have computed the correlation function 
between flight lengths $\langle \Delta v(t) \Delta v(0) \rangle$ and found that it 
falls only very slowly, perhaps as $t^{-\kappa}$ with the exponent around 
$\kappa\approx 0.1$. 
\par
We note again the fact that the map (\ref{uvmap}) exhibits anomalous 
diffusion also for a pseudointegrable case of {\em rational} $\mu$, 
and irrational $y_0$. Irrational $y_0$ can be successively approximated by rational 
initial momenta $y_0 \approx 2 m_i/l_i$, e.g. by its continued fraction approximations, for 
which the orbit is ballistic. Since the system is parabolic, such approximate orbit will 
follow an exact orbit up to times of order $t \sim l_i$, which is typically smaller than the 
period of ballistic orbit compactified on a torus $\sim l_i^2$.
Hence there is no contradiction between the ballistic spreading for the rational 
approximates and the anomalous spreading for the irrational initial conditions. 
The question of explaining anomalous diffusion can perhaps be reduced to the 
number theoretic study of periodic orbits.  
Note also that anomalous diffusion implies fractal singular continous spectrum of
classical time evolution\cite{Artuso00}.
\par
We now turn to the quantum version of the map (\ref{uvmap}). Quantization goes 
along the standard lines (e.g. \cite{Izrailev}), writing the Hamiltonian of the map 
(\ref{uvmap}) as $\op{H}=\op{T}(\op{v}) + \delta_{\rm p}(t)V(\op{u})$, where 
$\delta_{\rm p}(t)=\sum_n{\delta(t-n)}$, the kinetic part is 
$\op{T}(\op{v})=\op{v}^2=-\partial^2/\partial u^2$ and the potential is
\begin{equation}
V(u)= (|u|+\mu-1/2)^2, \quad {\rm for} \;\; |u|\le 1,\quad
\label{potential}
\end{equation}
and is periodic in $u$ with the period 2. 
Here we shifted the origin of $u$ to the $1/2$ with respect to
(\ref{uvmap}) in order to have a symmetry $V(u)=V(-u)$. Note that potential 
$V(u)$ has a discontinuous derivative at integer values of $u$.
In order to have a finite Hilbert space for numerical work, 
we compactify our cylindrical phase space to the torus 
$(u\hbox{ (mod }Q),v\hbox{ (mod }P))$, where $Q\equiv 2$ and $P$ is the periodicity 
(number of fundamental cells) in $v$-direction. Because we want to study diffusion, 
periodicity in the momentum $P$ will be set to some large integer, so that the phase 
space will effectively mimic a cylinder. Using periodic boundary conditions 
$\langle u+Q| \psi \rangle = \langle u |\psi \rangle$,
$\langle v+P | \psi \rangle = \langle v| \psi \rangle$, in position, momentum representation,
respectively, $\hbar$ must satisfy $QP/2\pi \hbar = N$, where $N$ 
is an integer giving the dimension of the Hilbert space. 
There are natural finite position and momentum bases satisfying the above boundary 
conditions, namely $\ket{u_j}$ and $\ket{v_k}$ respectively, with 
$u_j=(j-N/2)Q/N,j=0,1,\ldots N-1$ and 
$v_k=(k-N/2)P/N,k=0,1,\ldots N-1$. 
We perform quantum evolution $\ket{\psi(t)} = \op{U}^t \ket{\psi(0)}$
by decomposing the unitary propagator $\op{U} = \exp(-i\op{T}/\hbar)\exp(-i\op{V}/\hbar)$ 
to {\em diagonal} kinetic and potential part, with entries 
$\exp(-i v_k^2/\hbar)$ and $\exp(-i V(u_j)/\hbar)$ 
respectively, where Fourier transformation between the two bases 
$\braket{u_j}{v_k} = N^{-1/2} {\rm e}^{{\rm i}u_j v_k/\hbar}$ is coded efficiently
using FFT-algorithm. 
Our main goal was to simulate time evolution and check quantum-classical correspondence with 
anomalous diffusion for short times and possibility of dynamical localization for longer times.
We argued earlier that local (in)stability of a classical system may not play a 
key role in dynamical localization. We know that in cases of normal classical 
diffusion one often finds pure point spectrum of quantum evolution, like e.g. in
kicked rotor \cite{Izrailev}, and that ballistic classical spreading, since it
is typically associated with regular structures in classical phase space, implies
(at least partly) continuous quantum spectrum.
However, the case of anomalous classical diffusion is much less clear. 
At least there are no examples with an anomalous classical diffusion, that 
would exhibit dynamical localization, i.e. pure point quantum spectrum. 
It is therefore important to check the 
nature of a quantum spectrum for our map. We have done this along two independent 
lines: direct time evolution in the momentum space and the 
solution of an eigenvalue problem. 
\begin{figure}
\centerline{\psfig{figure=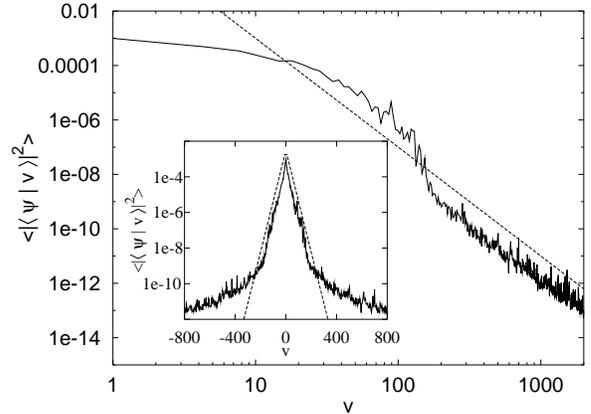,width=2.25in,angle=-90}}
\caption{Localized state after $t=10000$ iterations averaged over $\Delta t=200$ successive 
iterations, $(1/\Delta t)\sum_{t}^{t+\Delta t}|\langle v | \psi(t) \rangle |^2$, 
is shown. Wave function (full curve) has a power-law $\propto v^{-4}$ tail (dashed curve). 
In the inset is in semi-log scale drawn only central, exponentially localized part of a 
wave function. Dashed curve is an exponential function. Everything is for $P/N=6765/349196$ 
($M=50$).}
\label{fig:state}
\end{figure}    
\par
Following the time evolution of an initial momentum state 
$\langle v| \psi(0) \rangle:=\delta(v)$, 
we can check whether we have pure point spectrum or not. 
This has been performed using two different methods. 
The first method, which we call {\em torus approximation} (TA) has been described above 
and is a standard approach for dealing with quantum maps on cylinders.
We have chosen $P/N$ to be a continued fraction approximation
of a {\em noble} irrational $\pi \hbar = 1/(M+\sigma_{\rm GM})$, 
where $\sigma_{\rm GM}=(\sqrt{5}+1)/2$ and $M$ is an integer which determines 
the size of inverse $\hbar$. By increasing $M$ we approach the semiclassical limit
while increasing $P$ improves the accuracy of TA.
However, we should note that within TA 
due to periodicity $P$ in momentum direction $v$ 
we face {\em significant aliasing} of power-law tails in Fourier transform
of a {\em non-smooth} potential $V(u)$.
Thus we have devised an alternative quantum 
evolution with an {\em absorbing boundary} (AB) in the momentum space.
We truncate the Hilbert space to the set of $N+1$ momenta 
$v_k=\pi \hbar k$ with $k=-N/2,\ldots,N/2$, where now we can take any irrational value of 
$\hbar$ since we do not require periodicity in momentum space. 
The propagator for AB method is a 
finite {\em nearly-unitary} matrix 
$U_{k,l} = \theta(k)\exp(-i v^2_k/\hbar))\xi_{k-l}$ 
where $\theta(k)=1$ if $|k| \le N/2$ and $\theta(k)=0$ otherwise. 
Now we calculate the Fourier components of the potential propagator 
$\xi_k = (1/2)\int_{-1}^{1} du \exp(i\pi k u - iV(u)/\hbar)$ exactly in terms of complex
error function, that is with {\em no} aliasing, however we may loose some quantum probability 
leaking from the system for large momenta. Controlling the stability of the results, e.g. the 
quantum dispersion $\sigma^2(t) = \bra{\psi(t)}\op{v}^2\ket{\psi(t)}$
and the stay probability $\eta(t)=\braket{\psi(t)}{\psi(t)}$ with 
increasing $N$ we observe that AB method is typically much more efficient than TA, i.e. it 
converges for smaller dimension $N$. In general we found that results on $\sigma^2(t)$ of both 
methods agree up to some time scale which monotonically increases with increasing $N$. Beyond 
that time, $\sigma^2(t)$ of TA tends to overshoot due to aliased probability at large momenta, 
while $\sigma^2(t)$ of AB remains stable. 
\par
\begin{figure}
\centerline{\psfig{figure=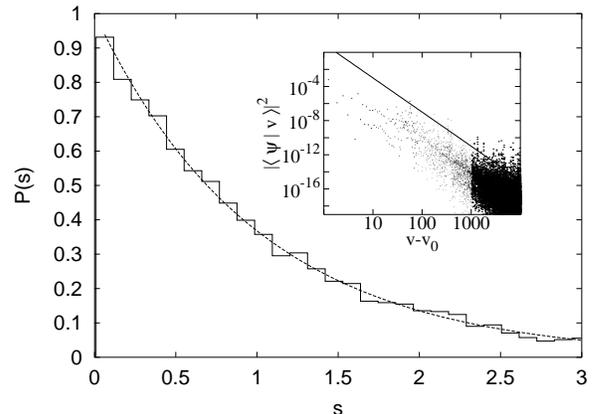,width=2.25in,angle=-90}}
\caption{Level spacing distribution $P(s)$ (full curve) for $M=0$ and $P/N=6765/10946$. 
Dashed curve is a Poisson distribution. In the inset we show a typical localized eigenstate 
of $\op{U}$ for $P/N=28657/46368$ ($M=0$)
peaked at $v_0=10858.9$ and the straight line gives the slope $-4$.}
\label{fig:poisson}
\end{figure}

In fig.\ref{fig:localization} time dependence of 
a quantum dispersion $\sigma^2(t)$ is shown for several values of $M$ together with the 
classical case. For cases $M=100$, $M=50$, and $M=0$ (inset) TA results are shown for
$P/N=6765/687446, 4181/215815$, and $P/N=121393/196418$, respectively, which do not deviate 
significantly from the AB results within entire time range $t < 10^5$.
However, for $M=25$ AB result for $N=60000$ satisfying $\eta(t) > 1 - 6\cdot 10^{-7}$ is 
shown instead, since we were not able to reach convergence of TA in the entire time range 
within accessible values of $N$.
One can see that the quantum evolution follows the classical diffusion up to some time, 
after which dynamical localization sets in. The localization is clearly demonstrated for small 
$M$, but for smaller $\hbar$ (e.g. $M=100$) the localization length seems to increase very 
fast with $M$ and the saturation could not be reached numerically within a reasonable time of 
computation. However, using standard heuristic arguments\cite{ChirikovCasati}, namely that
the quantum spread roughly follows the classical $\sigma^2 \sim t^\alpha$ until 
the relaxation (break) time $t=t_{\rm B}$ which is proportional to the density of 
`operative eigenstates' $t_{\rm B}\sim \hbar^{-1} \sigma$, we derive an
asymptotic $\hbar$-dependence of the localization length,
$ \sigma_{\rm l}(\hbar) \propto 1/\hbar^{\alpha/(2-\alpha)}. $
In our case, since $\alpha$ is quite close to ballistic value of $2$ giving formally 
$\sigma_{\rm l}=\infty$, the localization length is predicted to grow very fast asymptotically,
$\sigma_{\rm l} \propto \hbar^{-9.5} \propto M^{9.5}$.
Of course, we should be aware that the asymptotic scaling $\sigma_{\rm l}(\hbar)$
is effectively not very accurate for not very small $\hbar$ and not very long times 
$t$ where the classical spread increases with a smaller and varying effective exponent.
In addition, 
localized states of time evolution $\langle v_k|\psi(\infty)\rangle$, as well
as eigenstates of $\op{U}$, have an interesting 
structure: an initial exponential decay with a rapid crossover to an asymptotic 
power law tail ($\sim v^{-4}$) which is expected\cite{Mirlin} due to non-smooth $C^0$ 
potential (\ref{potential}) (see fig. \ref{fig:state}). 
For smaller $M$ the width of the central exponentially localized part of the wave function is 
smaller and the main contribution to $\sigma^2$ comes from the spiky fluctuating power-law 
tails, as a results of which $\sigma^2(t)$ exhibits big fluctuations as a function of time
(see insets in figs.\ref{fig:localization},\ref{fig:poisson}).
On the other hand, in the semiclassical limit the central exponential part 
of the wave function becomes wider and consequently the tails become less important.
\par
The results of a stationary analysis of quasi-energy spectrum and eigenstates
of the quantum propagator are fully compatible with the results of direct time evolution.
We have diagonalized the matrix ${\rm U}_{k,l}$ within TA and all eigenfunctions were found to
be localized. We also analyzed the spectral nearest neighbor spacing distribution
$P(s)$ which, as expected, is always found to have a Poisson shape $\exp(-s)$ 
(see fig. \ref{fig:poisson}). 
We must remark that by decreasing the size of the Hilbert space $N$ at fixed $\hbar$ (or $M$),
we obtain a transition to random matrix (Wigner-like) spectral 
distribution when $P$ becomes comparable or smaller than localization length 
$\sigma_l$.  
\par
In conclusion, we have demonstrated dynamical localization in a quantum system whose 
classical counterpart is a non-integrable, piecewise parabolic (locally stable) map and 
exhibits super-normal anomalous diffusion. The result implies that local dynamical 
instability of a classical system is not necessary for dynamical localization. 
The mechanism of anomalous diffusion is here more complicated than in hierarchical chaotic 
systems, as the consecutive L\' evy flights are strongly correlated. 

Financial support by the Ministry of Education, Science and Sports of Slovenia is
gratefully acknowledged.

\end{document}